\documentclass[twocolumn,aps,showpacs,floatfix]{revtex4}
\usepackage{graphicx}
\usepackage{dcolumn}
\usepackage{amsmath}
\usepackage{natbib}

\begin{document}
\title{Rhythmogenic neuronal networks, pacemakers, and k-cores}
\author{David J. Schwab}
\author{Robijn F. Bruinsma}
\affiliation{Department of Physics and Astronomy, University of California, Los Angeles, CA 90095-1596}
\author{Alex J. Levine}
\affiliation{Department of Chemistry \& Biochemistry and The California Nanosystems Institute\\
UCLA, Los Angeles, CA 90095-1596}

\date{\today}

\begin{abstract}
Neuronal networks are controlled by a combination of the dynamics of
individual neurons and the connectivity of the network that links
them together. We study a minimal model of the \emph{preBotzinger
complex}, a small neuronal network that controls the breathing
rhythm of mammals through periodic firing bursts. We show that the
properties of a such a randomly connected network of identical
excitatory neurons are fundamentally different from those of
uniformly connected neuronal networks as described by mean-field
theory. We show that (i) the connectivity properties of the networks
determines the location of \emph{emergent pacemakers} that trigger
the firing bursts and (ii) that the collective desensitization that
terminates the firing bursts is determined again by the network
connectivity, through \emph{k-core clusters} of neurons.
\end{abstract}

\maketitle

A neuronal network is a group of interconnected neurons functioning
as a circuit. Each neuron receives electrical signals from a
collection of tree-like dendrites, connected via synaptic junctions
to the branched output terminals of other neurons. The neuron
responds, based on some function of its input signals, by either
doing nothing or by ``firing,'' i.e., by producing an
action-potential output pulse that is received by other
neurons~\cite{Nelson:03}. In a network of excitatory ``integrating''
neurons, the electrical potential of the cell body of a neuron
always increases by an amount $\Delta V$ when the cell receives a
voltage input pulse. The potential of the cell effectively
integrates the signals from other neuronal outputs. The firing
probability of a neuron depends sensitively on the electrical
potential of the cell, leading to threshold behavior in which the
neuron can be considered to be either in a quiescent state
characterized by sporadic firing if the cell potential is large and
negative (``hyperpolarized''), or an activated state at higher
potentials (``depolarized''), with more than an order of magnitude
increase in firing rate over the quiescent state.

A classical example of an integrating neuronal network is the
preB\"{o}tzinger Complex (pBC) of about $10^2$ neurons located in
the brain stem~\cite{Feldman:90,Smith:91}. In this network, which
collectively produces a rhythmic voltage signal that sets the timing
of inspiration in mammals under resting conditions, each neuron is
connected on average to one-sixth of the other neurons. The period
of the current bursts is on the order of a second, which is about
$10^2$ times longer than the time scale associated with repeated
firing by activated individual neurons. The slow modulation is
believed to be due to calcium-mediated ``adaptation.'' With each
input pulse, the dentritic calcium concentration increases by an
amount $\Delta C$. The increase in calcium concentration leads to an
increase in the leakage conductance between the dendrites and the
surrounding medium, making the neuron's somatic potential
insensitive to incoming action potentials. When the somatic neuron
potential drops below the threshold, it stops firing. After a
recovery period during which the dendritic calcium relaxes back to
its equilibrium value, the neuron once again begins integrating input signals.

Two different mechanisms have been proposed to explain the
synchronization of the firing of the different neurons of the pBC.
Neurons that in isolation can oscillate autonomously between cycles
of firing and quiescence are known as
\emph{pacemakers}~\cite{Butera:99}. In the individual pacemaker
hypothesis, it is assumed that the pBC neurons are slaved to a small
number of pacemakers believed to be present in the pBC. In the
emergent pacemaker hypothesis (EPH), it is assumed that the
oscillation is a collective property of a large group of neurons
that would not oscillate in isolation~\cite{DelNegro:08}. True
pacemaker neurons, if present, only provide a back-up function and
are not essential for the oscillation. The oscillations are in this
case expected to disappear if the number of neurons drops below some
threshold value. Indeed, when more than $80\%$ of these pBC neurons
are destroyed in an \emph{in vitro} experiment, the firing sequence
changes from periodic oscillation to an increasingly complex
pattern. This also happens when the excitability of the neurons is
increased~\cite{DelNegro:02}.

In this letter we show that the ability of a non-uniform neuronal
network to collectively generate an oscillation, as assumed in the
EPH, depends on general properties of the network connectivity,
independent of the details of the model of neuron dynamics;
moreover, these properties can be analyzed in terms of simple
graph-theoretic methods~\cite{Newman:03,Timme:06}. Specifically, we
show that: (i) a network of identical but randomly connected neurons
supports periodic synchronized bursts triggered by those neurons
that are linked to a maximum number of other network neurons through
a minimum number of links, and (ii) for highly excitable networks,
the minimum number of neurons required for rhythmogenesis is
determined a sequence of {\em Magical Numbers}, which are a function
solely of the {\em adjacency matrix.}

We demonstrate these claims using a simple model for an excitatory
neuronal network regulated by calcium-based adaptation. Each neuron
is represented by two dynamical variables: the somatic potential
$V_i(t)$  and the calcium concentration  $C_i(t)$ of the $i^{\rm
th}$ neuron. The $N$ neurons fire according to the $2N$ coupled
non-linear rate equations:
\begin{eqnarray}
\label{V-eqn}
\frac{d V_i}{dt} &=& \frac{1}{\tau_V} \left( V_{\rm eq} - V_i \right) +
\Delta V(C_i) \sum_{j \ne i}M_{ij} P(V_j) \\
\frac{d C_i}{dt} &=& \frac{1}{\tau_C} \left( C_{\rm eq} - C_i \right) +
\Delta C \sum_{j \ne i}M_{ij} P(V_j)
\label{C_eqn}
\end{eqnarray}
$V_{\rm eq}$ and $C_{\rm eq}$ are, respectively, the resting
potential and the equilibrium calcium concentration of a neuron with
$\tau_V$ ($10$ms) and $\tau_C$ ($500$ms)~\cite{Koch:99} the
respective equilibration times (the calcium concentration is thus
the slow variable). Calcium-mediated adaptation is allowed for by
assuming that $\Delta V(C)$ drops rapidly when the calcium
concentration $C$ exceeds a threshold $C^*$. The time-sequence of
firing events generated by a neuron is assumed to be a Poisson
process with a voltage-dependent mean firing rate $P(V)$. For $P(V)$
we will assume that if $V$ exceeds the threshold $V^*$ then $P(V)$
increases from a basal rate of about five spikes per second to a
high rate of about seventy-five spikes per second. Finally, the
entries of the adjacency matrix $M_{ij}$  are equal to one if the
output of $j^{\rm th}$ neuron is an input to neuron $i$, and zero
otherwise.

\begin{figure}
\includegraphics[width=6cm]{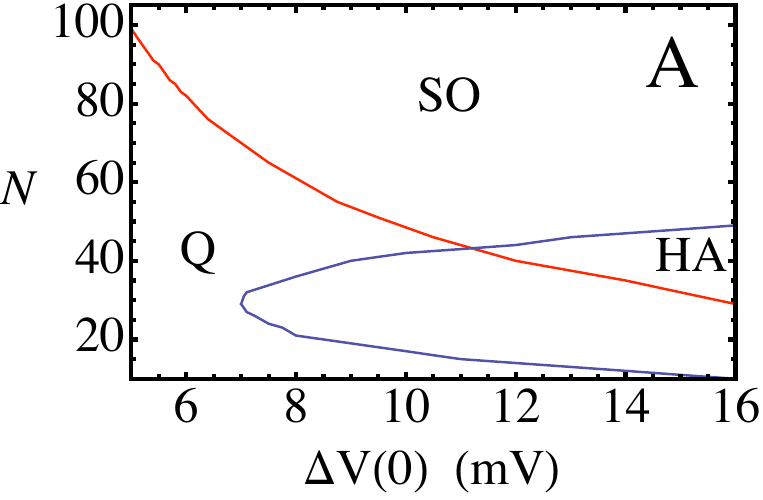}
\includegraphics[width=6cm]{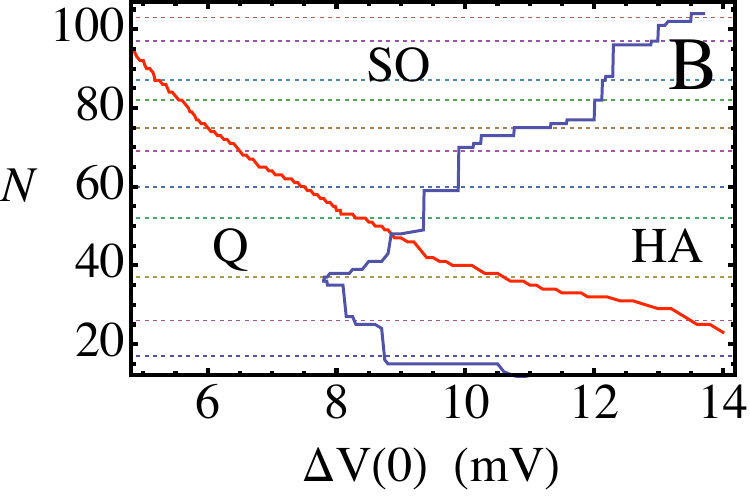}
\caption{(color online) (A) Phase diagram of a homogeneous
$N$-neuron network with every neuron linked to every other neuron.
The horizontal axis is the maximum voltage jump of a cell following
an action-potential input pulse. The red line is the stability limit
of a low-activity fixed point of
Eqs.~(\ref{V-eqn-MFT}),(\ref{C_eqn-MFT}) (Q phase) while the blue
line is the stability limit of a high-activity fixed point (HA
phase). Cooperative limit-cycle oscillations are fully stable only
in the region above the blue and red lines (SO phase). (B) Phase
diagram of an inhomogeneous, random $N$-neuron network with, on
average, each neuron linked to $N/6$ other neurons. In the section
labeled HA deterministic chaos, period-doubling and intermittency is
encountered. The dashed lines mark a sequence of Magical Numbers
$N_k$ determined by the adjacency matrix of the network.}
\label{phase-diagram-pic}
\end{figure}

We start with the very simplest case of a homogeneous network where
every neuron is linked to every other neuron in both directions:
$M_{ij}=1$ for all $i,j$ (known as a ``clique''). If the initial
potentials and calcium concentration also are the same for all
neurons, then the $2N$ rate equations reduce to a single pair that
describes all neurons:
\begin{eqnarray}
\label{V-eqn-MFT}
\frac{d V}{dt} &=& \frac{1}{\tau_V} \left( V_{\rm eq} - V \right) +
N \Delta V(C)  P(V) \\
\frac{d C}{dt} &=& \frac{1}{\tau_C} \left( C_{\rm eq} - C \right) +
N \Delta C P(V), \label{C_eqn-MFT}
\end{eqnarray}
which can be analyzed by the standard methods of dynamical
systems~\cite{Izhikevich:07}. The pair of equations can also be
viewed as a ``mean-field'' approximation for more complex
networks~\cite{Bressloff:00}. The resulting dynamical phase diagram
is shown in the upper panel of Fig.~\ref{phase-diagram-pic}.

In the parameter regime marked SO (``stable oscillation''), the
potential and calcium concentrations of the neurons undergo a stable
limit-cycle oscillation. For lower values of the input voltage jump
at zero calcium concentration $\Delta V(C=0)$, corresponding to
weakly excitable neurons, the period of the oscillation increases
and then diverges as the number N of neurons is reduced due to the
appearance of an infinite-period saddle-node
bifurcation~\cite{Ermentrout:94}. A line of these bifurcations
separates the SO phase from a quiescent phase, marked Q, where all
neurons are permanently in a state of low activity. For higher
values of $\Delta V(C=0)$, corresponding to highly excitable
neurons, the unstable fixed point at the center of the limit cycle
becomes stable as N is reduced. In the part of the phase diagram
where this happens, marked HA, the neurons are permanently in a
state of high activity. This mean-field HA phase does {\em not} show
the complex firing pattern reported experimentally when the
excitability was increased~\cite{DelNegro:02}.

In actuality, each neuron of the pBC is believed to be linked to
$(1/6)^{\rm th}$ of the other neurons so the pBC network is not a
clique. To describe this, we use an Erd\H{o}s-R\'{e}nyi random
adjacency matrix~\cite{Solomonoff:51, Erdos:59}, assigning zeros and
ones as the entries of $M_{ij}$ with probabilities $5/6$ and $1/6$,
respectively. Solution of the coupled rate equations on a single random
graph produces the phase-diagram shown in the lower panel of
Fig.~\ref{phase-diagram-pic}.  The heterogeneity of the network does
not destroy its ability to produce robust, synchronized stable
oscillations, though note that the SO section of the phase diagram
has been reduced in area as compared to the mean-field case.

\begin{figure}
\includegraphics[width=6cm]{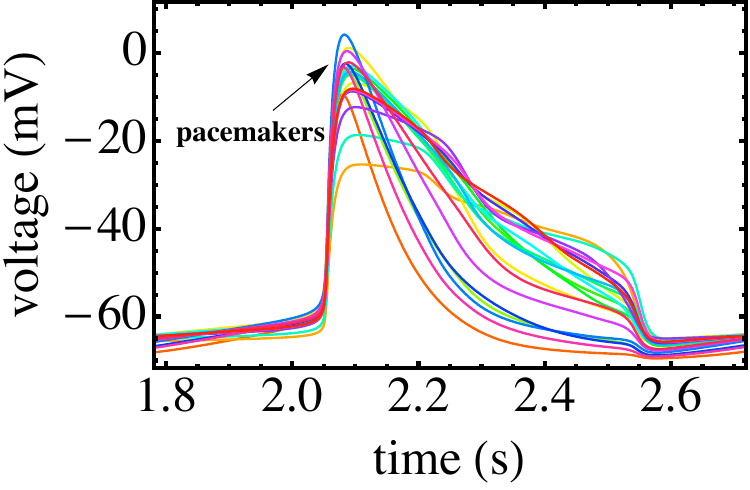}
\includegraphics[width=6cm]{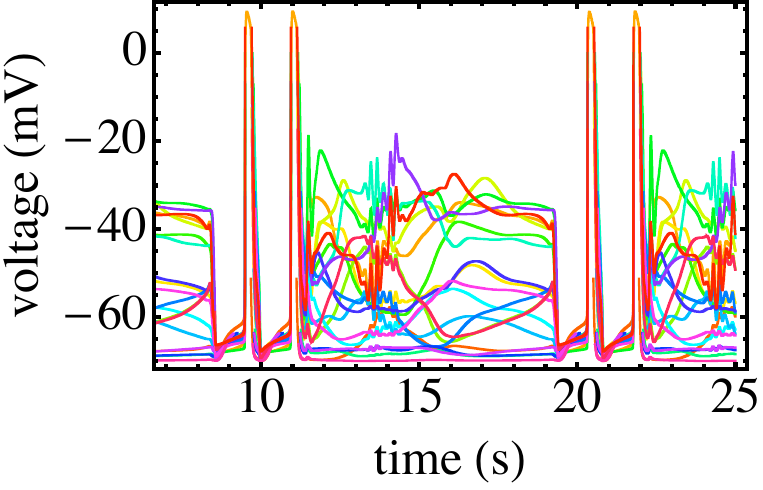}
\caption{(color online) (A) Periodic potential oscillations (SO
phase) for a random connectivity matrix. Different colors refer to
the different neurons. In the low-activity part of the cycle, the
potential of all neurons is below the firing threshold ($-55 $mV).
The potential of  a limited number of neurons (e.g. blue) rises
significantly more quickly at the initiation of a burst. When one of
these ``pacemaker'' neurons crosses the threshold, it triggers a
voltage avalanche that spreads over the whole network. (B)
Time-dependent potentials in the high-activity phase. Multiple
collective potential bursts alternate with an incoherent, chaotic
state. Note the different time scales in panels A and B.}
\label{spikes-pic}
\end{figure}

Unlike the mean-field case, the firing pattern now varies greatly
from one neuron to the next. Superimposing the firing patterns of
different neurons reveals an important feature (see
Fig.~\ref{spikes-pic}A).
In the low-activity part of the cycle, the potentials of all neurons
are below $V^*$, but they rise more rapidly for a sub-population; these
reach the firing threshold first. Their increased firing rate pushes
sub-threshold neurons linked to them past the firing threshold as
well. A chain reaction spreads over the network until all neurons
are above threshold. Note that the least excitable neurons that
crossed the firing threshold latest remain active over a longer
period of time, producing a highly asymmetric pulse shape. Even
though in our model all neurons are identical, and none can oscillate
autonomously, a few neurons, selected through the network
connectivity, are timing the oscillations. This subpopulation of
spike leaders can be interpreted as the emergent pacemakers of the
network. The other neurons effectively amplify their action.
\begin{figure}
\includegraphics[width=6.0cm]{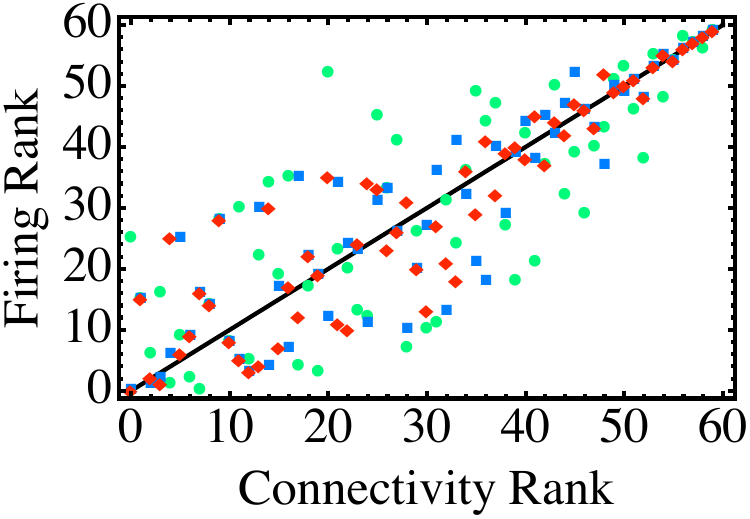}
\caption{(color online) Predicted rank: neurons ranked according to
the number of neurons linked to them by one input link (green
circles, ``Level 1''), one or two input links (blue squares, ``Level
2''), one, two or three input links (red diamonds, ``Level 3'').
Actual rank: neurons ranked according to their somatic potential
preceding a firing burst. Low rank corresponds to high potential.
The neuron with lowest predicted rank indeed is the neuron with the
lowest actual rank, but predicted rank correlates best with actual
rank for higher rankings} \label{pacemakers-pic}
\end{figure}

Network degradation, i.e., randomly knocking out neurons, leads to
complex changes in the set of these emergent pacemaker neurons. It
also takes longer for them to reach threshold as N is reduced
so the oscillation period increases. For lower values of  $\Delta
V(C=0)$, i.e., weakly excitable networks, the period diverges along
the phase-boundary between the SO and Q phases in
Fig.~\ref{phase-diagram-pic}B, which agrees with the predictions of
mean-field theory. For higher values of $\Delta V(C=0)$ the system
enters the HA phase. Unlike in mean-field theory, the HA phase of
the random network exhibits the complex dynamical behavior, with
period doubling and deterministic chaos, that was reported
experimentally~\cite{DelNegro:02}. One example of this is shown in
Fig.~\ref{spikes-pic}B where groups of high-activity bursts
alternate with periods of deterministic chaos, forming a complex limit cycle.
\begin{figure}
\includegraphics[width=7cm]{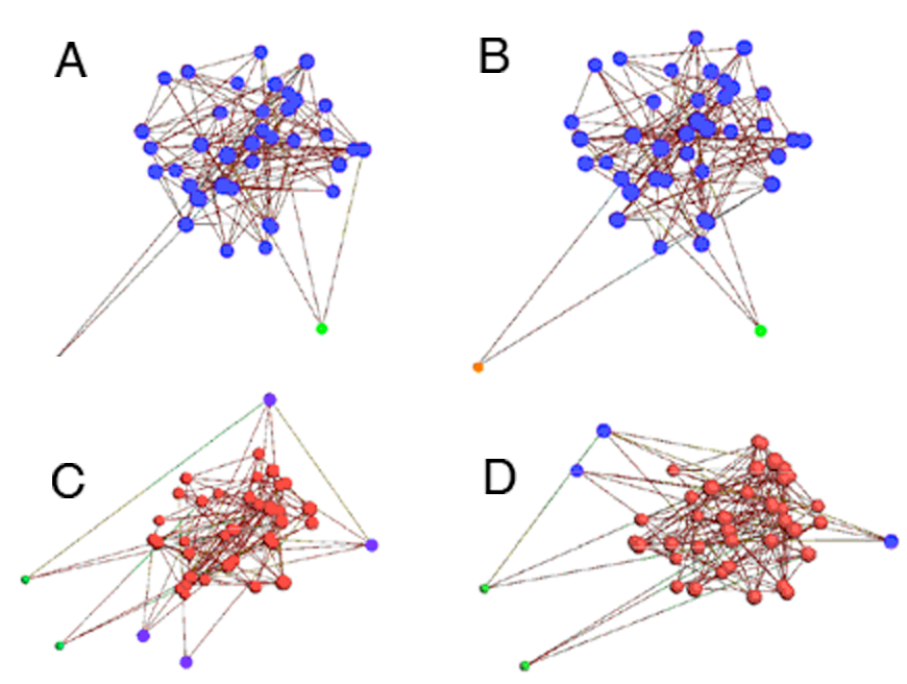}
\caption{(color online) $k$-cores of a symmetric $N \times N$ random
adjacency matrix. Nodes making up the $5$-core are marked in red,
$4$-core nodes in blue, $3$-core nodes in green, and $2$-core nodes
in orange. The radial distance of a node from the center increases
with decreasing $k$. With a given $k$-core, a node's radial position
is increased by connections to lower $k$-core nodes and decreased by
connections to higher $k$-core nodes. The four figures show a
progressively increasing network size: ($N_A=40$, $N_B=41$,
$N_C=42$, $N_D=43$). Image created using
Ref.~\cite{Alvarez-Hamelin}} \label{k-core-pic}
\end{figure}

The $\mbox{SO} \leftrightarrow \mbox{HA}$ threshold curve
$N_c(\Delta V)$ has a surprising stair-case dependence on $\Delta V$
differing dramatically from the continuous curve predicted by the
mean-field theory (see Fig.~\ref{phase-diagram-pic}). These
discontinuities define certain privileged numbers of neurons $N_k$
for which the network fails to support stable oscillations. The
values of these $N_k$'s are independent of system parameters such as
$\Delta C$.  The boundary of the SO regime makes discrete jumps
between $N_k$'s  as the parameters ($\Delta C$, $\Delta V$) of the
neuronal dynamical model are changed. In contrast, the phase
boundary separating the SO and Q phases in
Fig.~\ref{phase-diagram-pic}B largely follows mean-field
predictions.

Both the selection of the pacemaker neurons and the values of the
privileged number of neurons appear to be determined largely by the
mathematical properties of the adjacency matrix $M_{ij}$ independent
of the details of our dynamical model. In Fig.~\ref{pacemakers-pic}
we show how one can identify the pacemaker neurons through their
connectivity by ranking them according to the number of neurons
connected to a given neuron by no more than three links. These
pacemaker neurons do not, however, play a central role in the
determination of the minimum number of network neurons able to
support collective oscillations for highly excitable neuronal
networks. While sufficiently active synaptic inputs to a pacemaker
indeed will more quickly drive the dendritic calcium concentration
past the threshold $C^*$, so that it becomes desensitized and thus
unable to spike, a small set of such desensitized and quiescent
neurons will not drive an active network to collectively
desensitize. Rather, one must have a system-spanning
high-connectivity network capable of simultaneously desensitizing
\emph{all} of the neurons to quiet the inherently excitable system.

To quantify the size of a high-connectivity part of the network, it
is useful to introduce the concept of a
$k$-core~\cite{Dorogovtsev:06}. A $k$-core of a graph (for integer
$k$) is a subgraph in which all nodes (i.e. neurons) have at least
$k$ inputs from other nodes in the subgraph. As the number of nodes
increases in an Erd\H{o}s-R\'{e}nyi random network, $k$-core
clusters appear with larger $k$ values at sharply defined
thresholds. As an example we show in Fig.~\ref{k-core-pic}A the
$k$-cores of a symmetric $40 \times 40$ random adjacency matrix.
Nearly all nodes form a single $4$-core cluster. Adding one more
node at random does not change this feature
(Fig.~\ref{k-core-pic}B), but adding two nodes at random, so that $N
= N_5 = 42$ produces a sharp transition in which the network is now
dominated by a single, system-sized $5$-core cluster
(Fig.~\ref{k-core-pic}C). Adding an additional node to $N=43$ does
not alter the dominance of the 5-core, as shown by
Fig.~\ref{k-core-pic}D. For the random network used to generate
Fig.~\ref{phase-diagram-pic}B, these discontinuous transitions take
place at $N_3 = 17$, when a $3$-core appears; at $N_4 = 26$ when a
$4$-core appears; at $N_5 = 37$ when a $5$-core appears, and so on.
The values of $N_k$  for this realization of the random network are
represented by dotted lines in Fig.~\ref{phase-diagram-pic}B. The
locations of the discontinuities of the phase boundary as a function
of $N$ agree well, though not perfectly, with the $k$-core
transition values $N_k$. The discrepancies are presumably due to the
fact that a member of a $k$-core can have more than $k$ input links,
including links from non $k$-core neurons. We emphasize that the
$k$-core concept is inapplicable to the $\mbox{SO} \leftrightarrow
\mbox{Q}$ transition. Along the transition line, the few neurons
with the highest connectivity are able to trigger an excitation wave
that spreads through the whole system. These few emergent pacemakers
are simply outliers having maximal connectivity and need not be part
of a high $k$ $k$-core.

In summary, we have presented a simple model for rhythmogenic
neuronal networks, such as the pBC, using a combination of excitable
integrate-and-fire neurons modified by a slower process of
calcium-mediated desensitization. The most important conclusion of
our work is that key features of the network dynamics -
determination of the pacemaker neurons and determination of the
minimal number of neurons that support stable oscillation - are
determined (largely) by network connectivity. We also showed that in
the phase diagram there is an asymmetry between the transition from
the stable oscillation phase to the quiescent phase and the
transition from the stable oscillation phase to the high activity
phase. The first transition is well described by mean-field theory,
while the staircase structure of the phase boundary of the second
transition reflects the full nature of network connectivity. This
asymmetry originates from the difference between the dynamics of a
spreading wave of voltage-mediated excitation and collective,
calcium-mediated desensitization. Tests of the model should be
straightforward. The excitability of neurons can be increased in
experiment, effectively controlling the size of the action potential
$\Delta V$ in our model. Measuring the onset action potential for
complex firing patterns as a function of the number $N$ of neurons
should then directly reveal the predicted staircase structure of
Fig.~\ref{phase-diagram-pic}B.

We thank J. Feldman for enjoyable conservations and for
sharing unpublished data on pBC dynamics.

\end{document}